\documentclass[sigconf,nonacm]{acmart}
\usepackage{csquotes}
\usepackage{tcolorbox}
\usepackage{enumitem}
\AtBeginDocument{%
  }

\begin{document}

\title{From Questions to Trust Reports: A LLM-IR Framework for the TREC 2025 DRAGUN Track}



\author{Ignacy Alwasiak}
\email{ignacy.alwasiak@student.uj.edu.pl}
\affiliation{%
  \institution{Jagiellonian University}
  \city{Krakow}
  \country{Poland}
}

\author{Kene Nnolim}
\email{knnolim@mit.edu}
\affiliation{%
  \institution{Massachusetts Institute of Technology}
  \city{Cambridge}
  \state{MA}
  \country{USA}
}

\author{Jaclyn Thi}
\email{jthi@mit.edu}
\affiliation{%
  \institution{Massachusetts Institute of Technology}
  \city{Cambridge}
  \state{MA}
  \country{USA}
}

\author{Samy Ateia}
\email{samy.ateia@ur.de}
\affiliation{%
  \institution{University of Regensburg}
  \city{Regensburg}
  \country{Germany}
}

\author{Markus Bink}
\email{markus.bink@ur.de}
\affiliation{%
  \institution{University of Regensburg}
  \city{Regensburg}
  \country{Germany}
}

\author{Gregor Donabauer}
\email{gregor.donabauer@ur.de}
\affiliation{%
  \institution{University of Regensburg}
  \city{Regensburg}
  \country{Germany}
}

\author{David Elsweiler}
\email{david.elsweiler@ur.de}
\affiliation{%
  \institution{University of Regensburg}
  \city{Regensburg}
  \country{Germany}
}

\author{Udo Kruschwitz}
\email{udo.kruschwitz@ur.de}
\affiliation{%
  \institution{University of Regensburg}
  \city{Regensburg}
  \country{Germany}
}


\begin{abstract}
  The DRAGUN Track at TREC 2025 targets the growing need for effective support tools that help users evaluate the trustworthiness of online news. We describe the UR\_Trecking system submitted for both Task 1 (critical question generation) and Task 2 (retrieval-augmented trustworthiness reporting). Our approach combines LLM-based question generation with semantic filtering, diversity enforcement using clustering, and several query expansion strategies (including reasoning-based Chain-of-Thought expansion) to retrieve relevant evidence from the MS MARCO V2.1 segmented corpus. Retrieved documents are re-ranked using a monoT5 model and filtered using an LLM relevance judge together with a domain-level trustworthiness dataset. For Task 2, selected evidence is synthesized by an LLM into concise trustworthiness reports with citations. Results from the official evaluation indicate that Chain-of-Thought query expansion and re-ranking substantially improve both relevance and domain trust compared to baseline retrieval, while question-generation performance shows moderate quality with room for improvement. We conclude by outlining key challenges encountered and suggesting directions for enhancing robustness and trustworthiness assessment in future iterations of the system.
\end{abstract}

\keywords{Understanding News, Retrieval Augmented Generation, TREC}


\maketitle

\section{Introduction}
The proliferation of misinformation and disinformation across large-scale social networks and news outlets is a complex socio-technical phenomenon \cite{zhou2022confirmation, prochaska2023mobilizing, do2022infodemics}, with significant implications for how individuals form and adjust their opinions \cite{hasibuan2024role}. Existing interventions aim to mitigate these risks, for example, through algorithmic filtering \cite{collins2021trends} of inappropriate content, external fact-checking services \cite{hameleers2020misinformation}, or community-driven initiatives such as community notes \cite{prollochs2022community, godel2021moderating}. However, such measures are limited to specific platforms and contexts. Once misleading claims circulate beyond these controlled environments, users are left without institutional or algorithmic safeguards and must independently assess the credibility of the information they encounter.

Although many people believe they are good at discerning truth from fiction, studies show that users are often overconfident in their actual evaluation capabilities and thus ill-equipped \cite{mahmood2016people}. Similarly, search behavior can reinforce user-belief in misleading news \cite{aslett2024online,elsweiler2025query}. This highlights the urgent need to support users in their information search and evaluation practices.

The \textbf{DRAGUN} (Detection, Retrieval, and Augmented Generation for Understanding News) Track at TREC 2025 \cite{zhang2025dragun} addresses this challenge by proposing two related tasks: Task 1 focuses on generating critical and investigative questions that prompt readers to consider aspects such as source bias, underlying motivations, and diversity of viewpoints when evaluating online content. Task 2 complements this by generating detailed, retrieval-augmented (RAG-based) reports that help users thoughtfully assess the credibility of specific webpages.

Together, these question prompts and analytical reports aim to strengthen users' capacity to discern the trustworthiness of online information and make more informed judgments about the veracity of central claims in news articles.

As part of this paper, we present our participation in the two tasks by describing the resources and methodology we used along with the results we achieved in the official evaluation phase.

To support the reproducibility of our work, we publicly make available all implementations related to our submission on Github: \url{https://github.com/doGregor/UR_trecking_2025}.

\begin{figure*}[ht]
    \centering
    \includegraphics[width=\linewidth]{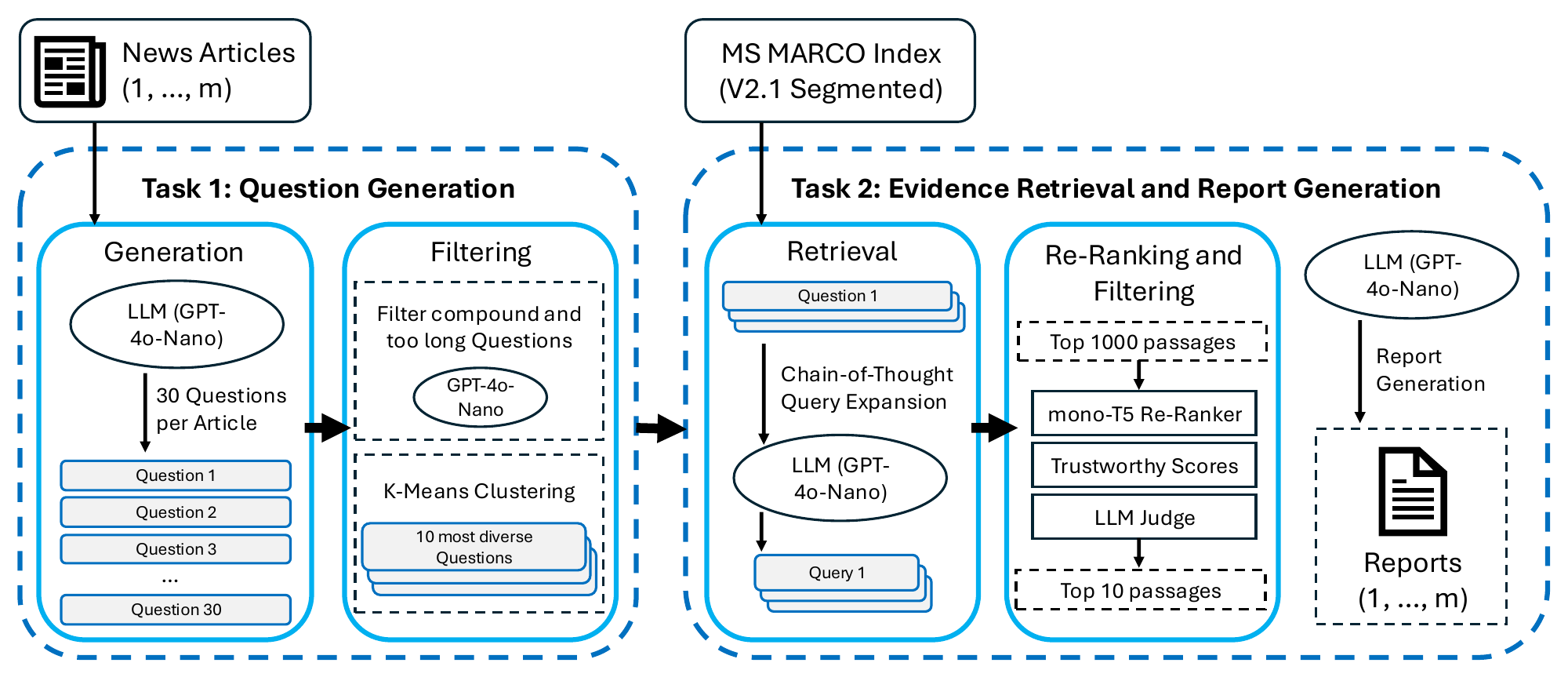}
    \caption{Flowchart of our pipeline for Tasks 1 and 2.}
    \label{fig:flowchart}
\end{figure*}

\section{Methods}

This section presents our end-to-end methodology (for an overview see Figure \ref{fig:flowchart}). We first describe the data, indexing setup, and language models used. We then outline our approach to generating and filtering critical questions, followed by retrieval with query expansion, re-ranking, and credibility-based filtering. Finally, we explain how the resulting evidence is synthesized into question-level answers and a final trustworthiness report.

\subsection{Data and Resources}

We indexed \textit{MS MARCO V2.1 (Segmented)}\cite{nguyen2016ms} in OpenSearch using the built-in default English analyzer. We use OpenAI's GPT4o-Nano for LLM-based components in our pipeline. This model was chosen primarily for its tradeoff between costs and efficiency.

\subsection{Question Generation and Filtering}
\label{subsec:qgen}

To generate questions, we used the aforementioned GPT model with a detailed prompt specifying the aspects we wanted the LLM to focus on. These included logical principles that the LLM should follow when generating questions and question formatting stipulations. For all details of this processing pipeline, including prompts and hyper-parameter settings, please consult the project repository\footnote{\url{https://github.com/doGregor/UR_trecking_2025}}.

The following is an excerpt of the prompt provided to the model:

\newtcolorbox{objective}{
  colback=gray!10,
  colframe=black,
  boxrule=1pt,
  arc=3pt,
  boxsep=0pt,
  left=10pt,
  right=10pt,
  top=10pt,
  bottom=10pt
}

\begin{objective}
\textbf{Follow these key principles for question generation:}

1) Investigate the Source:
\begin{itemize}[leftmargin=*]
    \item Generate questions about the publisher's (or key information sources such as organizations, experts, reporters, etc.) background, reputation, and potential biases.
    \item Ask about their credentials, expertise, and past work.
    \item Inquire about ownership, funding sources, and editorial policies.
\end{itemize}
...

\end{objective}


Once we had a set of questions generated by GPT, we applied semantic filtering to rid it of invalid questions; these included compound questions and questions that were too long. 

Finally, in order to ensure that our set of questions was diverse and covered a broad range of trustworthiness concerns, we applied unsupervised K-means Clustering on embeddings created for each question in our dataset via a sentence transformer. We then chose the question closest to each centroid (in embedding space) found after termination of the algorithm on each set of questions, divided by article, as the questions to keep for each provided article.

In order to evaluate the quality of the questions, we have calculated the TF-IDF score, Jaccard score, and cosine similarity between the question and the article in question. Additionally, we used a LLM to give scores to each question and check how well each question aligns with each CRAAP test\cite{blakeslee2004craap} component, indicating the reliability and credibility of sources based on currency, relevant, authority, accuracy and purpose. After manual inspection and calculations, we have found that there has been little correlation between the scores given by LLMs.


\subsection{Retrieval}
\label{sec:retrieval}

After generating critical questions for each target article, we issued these questions as queries to retrieve supporting evidence from the MS~MARCO~V2.1 segmented corpus. The purpose of this step was twofold: (1) to obtain candidate documents that could be used to answer each question, and (2) to collect citation-worthy evidence for use in the final trustworthiness report.


To improve retrieval quality beyond the baseline (using the original question as the query), we experimented with several query expansion strategies. Specifically, we evaluated the following methods:
\begin{itemize}
    \item \textbf{Baseline (No Expansion):} The original question was issued directly as a search query using BM25.
    
    \item \textbf{Boolean Expansion:} Additional keywords and phrases were generated and combined with the original query using Boolean operators (e.g., AND/OR) to broaden lexical coverage.
    
    \item \textbf{Chain-of-Thought (CoT) Expansion:} Following the approach described by Jagerman et al \cite{jagerman2023query}, we prompted a large language model to reason step-by-step about the information needed to answer a question and to generate expanded query terms accordingly.
    
    \item \textbf{Structured Expansion:} Instead of producing a single expanded query string, the language model generated a full OpenSearch Query DSL object. This allowed for more explicit control over field weighting and query composition.
\end{itemize}

To evaluate retrieval quality, we developed an LLM-based relevance judge to determine whether a retrieved document was judged as relevant to the given question. For each method, we examined the top 10 retrieved documents from the set of $k=1000$ candidates. A relevance score was computed as the proportion of these top documents that were labeled as relevant by the LLM.

\subsection{Re-ranking and Filtering}

We run multiple reranking and filtering steps on the list of retrieved documents.

First, we use the mono T5 reranker\footnote{\url{https://huggingface.co/castorini/monot5-base-msmarco}} \cite{nogueira-etal-2020-document} as fine-tuned on the MS~MARCO~V2.1 passage dataset to rerank the first 1000 retrieved documents.

We used the LLM-based relevance evaluator described in Section \ref{sec:retrieval} to determine whether each retrieved document meaningfully contributed to answering its corresponding question.

To assess the credibility of document sources, we incorporated the domain quality dataset introduced by Lin et al \cite{Lin2023DomainQuality}. It combines multiple standard datasets, such as NewsGuard, into a single comprehensive dataset. This dataset assigns a continuous trustworthiness score to news and web domains based on large-scale quality assessments.

For each retrieved document, we extracted its domain from the URL and looked up its corresponding trustworthiness score in the Lin et al.\ dataset. If a domain was not present in the dataset, it was assigned a default trust score of $0.0$. These domain scores were used as an additional signal for filtering and analysis.

After retrieval, query expansion, re-ranking, and relevance evaluation, we applied two final filtering strategies:

\begin{itemize}
    \item \textbf{Top-10 Relevant:} The top 10 documents labeled as relevant by the LLM.
    \item \textbf{Top-3 Relevant and Trustworthy:} The top 3 documents that were both labeled as relevant and had a domain trust score of at least $0.7$.
\end{itemize}

Filtering was applied over the top 100 documents produced after re-ranking. If fewer than the required number of documents satisfied the relevance or trustworthiness constraints, all qualifying relevant documents within the top 100 were retained. This final subset of documents was used for evidence synthesis and citation in the trustworthiness report.

This combined relevance-and-trust filtering process enabled us to focus on documents that were both semantically useful and originated from higher-quality information sources.

\subsection{Answer and Report Generation}

The \textbf{<Question, List of Related Articles>} tuples are analyzed by an LLM to create a sentence with explanation for each question that includes citations. After answering all of the questions an LLM shortens it into a report of reasonable length.

\section{Evaluation}

For evaluation, NIST assessors independently researched each news article's trustworthiness and created question-and-answer rubrics capturing what a good trustworthiness report should address. One primary assessor consolidated the three assessors' rubrics into a single rubric, which the DRAGUN organizers reviewed and for some cases lightly edited.

All assessor rubric question received an \textbf{importance level}: 
\begin{itemize}
    \item Have to Know (4 points)
    \item Good to Know (2 points)
    \item or Nice to Know (1 point)
\end{itemize}

To score the generated questions, two LLMs (Qwen3-Embedding-8B and Qwen3-Reranker-8B) selected the system question most similar to each rubric question. Human assessors then judged the selected pair(s) with one of four \textbf{similarity labels}: 

\begin{itemize}
    \item Very Similar (1 point)
    \item Similar (0.5 points)
    \item Different (0 points)
    \item or Very Different (0 points)
\end{itemize}

For each rubric question, the system earned the rubric question's importance weight multiplied by the highest similarity score available from its matched question(s). The final article-level score was the average of these weighted scores across all rubric questions.

\section{Results}

This section presents the results for Task~1 (question generation) and Task~2 (retrieval and report generation). We summarize the submitted run and analyze question quality, retrieval effectiveness under different query-expansion strategies, and the impact of re-ranking and trust filtering on final reports.

\subsection{Runs Submitted}


We submitted a single run for each task (Task 1 and Task 2). The submitted runs included all major components of our pipeline (LLM-based question generation, semantic filtering, clustering-based question selection, retrieval with multiple query-expansion strategies, re-ranking, and trust filtering). However, the iterative correction loop described in Section \ref{subsec:qgen} was \textbf{not} included in the final run due to time constraints. All remaining components were executed as described.

\subsection{Question Generation (Task 1)}

Across topics, our question-generation module produced between 10 and 20 candidate questions per article, depending on how many were removed through semantic and compound-question filtering. The official TREC evaluation assigned \textit{moderate} overall qgen\_scores, with substantial variation across topics (Table~1).

We consistently observed the following patterns:

\begin{itemize}
    \item \textbf{Very low contradictory scores}, indicating that our generated questions rarely conflicted with the assessor rubrics.
    \item \textbf{Low supportive scores}, suggesting that many questions did not strongly align with what human assessors considered most helpful for trustworthiness assessment.
    \item \textbf{High variance across topics}, with some articles scoring near zero and others showing noticeably stronger alignment (e.g., topics~35 and~41).
\end{itemize}

Overall, the results indicate that our pipeline effectively avoided generating harmful or contradictory questions, but struggled to consistently produce deeply supportive or rubric-aligned ones. This is likely due to the absence of the planned correction loop and instability in LLM-based scoring during development.

\subsection{Retrieval and Report Generation (Task 2)}

To evaluate retrieval quality, we compared all query-expansion approaches---baseline, Boolean, Chain-of-Thought (CoT), and Structured---before and after re-ranking. The results, visualized in Figures~1 and~2, show clear and consistent trends:

\begin{itemize}
    \item \textbf{CoT expansion achieved the highest relevance scores} across the first 20 TREC example questions and exhibited the \textbf{lowest variance} in performance.
    \item Boolean and Structured expansions displayed \textbf{higher variability}, suggesting greater sensitivity to prompt formulation and query composition.
    \item \textbf{Re-ranking consistently improved both relevance and domain trustworthiness} for all expansion strategies (Figures~3 and~4), demonstrating that the monoT5 reranker promotes not only semantically appropriate documents but also higher-quality domains.
\end{itemize}

After applying LLM-based relevance labeling and domain-trust filtering (threshold $\geq 0.7$), we generated the final RAG-based trustworthiness reports. Supportive scores for these reports were generally low (Table~1), reflecting that the retrieved evidence often addressed article claims only partially. Nonetheless, the stable improvements in both relevance and trust indicate that our retrieval pipeline meaningfully enhanced the quality of sources provided to the report-generation component.

In summary, the results show that \textbf{reasoning-based query expansion combined with re-ranking} delivered the most reliable retrieval performance, while question and report generation remained limited by components not fully integrated into the submitted pipeline.

\begin{table*}[h!]
\centering
\begin{tabular}{lllll}
\toprule
topic\_id & qgen\_score original & qgen\_score updated & contradictory\_score & supportive\_score \\
\midrule
msmarco\_v2.1\_doc\_04\_420132660 & 0.259 & 0.250 & 0.071 & 0.000 \\
msmarco\_v2.1\_doc\_06\_1440134319 & 0.350 & 0.333 & 0.000 & 0.000 \\
msmarco\_v2.1\_doc\_08\_300872161 & 0.158 & 0.100 & 0.000 & 0.075 \\
msmarco\_v2.1\_doc\_15\_116067546 & 0.200 & 0.192 & 0.051 & 0.038 \\
msmarco\_v2.1\_doc\_21\_861891150 & 0.226 & 0.219 & 0.031 & 0.312 \\
msmarco\_v2.1\_doc\_22\_1648697797 & 0.250 & 0.240 & 0.000 & 0.053 \\
msmarco\_v2.1\_doc\_25\_481628070 & 0.105 & 0.100 & 0.000 & 0.125 \\
msmarco\_v2.1\_doc\_25\_501708725 & 0.000 & 0.000 & 0.000 & 0.000 \\
msmarco\_v2.1\_doc\_25\_502424913 & 0.174 & 0.167 & 0.000 & 0.000 \\
msmarco\_v2.1\_doc\_34\_2529216 & 0.132 & 0.129 & 0.000 & 0.162 \\
msmarco\_v2.1\_doc\_34\_7751734 & 0.095 & 0.091 & 0.000 & 0.076 \\
msmarco\_v2.1\_doc\_35\_1300032609 & 0.450 & 0.429 & 0.000 & 0.114 \\
msmarco\_v2.1\_doc\_35\_441326441 & 0.133 & 0.125 & 0.000 & 0.000 \\
msmarco\_v2.1\_doc\_38\_227081897 & 0.353 & 0.111 & 0.000 & 0.000 \\
msmarco\_v2.1\_doc\_39\_1014551192 & 0.059 & 0.000 & 0.000 & 0.028 \\
msmarco\_v2.1\_doc\_39\_1165221603 & 0.077 & 0.074 & 0.049 & 0.019 \\
msmarco\_v2.1\_doc\_41\_1960853470 & 0.433 & 0.125 & 0.000 & 0.075 \\
msmarco\_v2.1\_doc\_42\_654618974 & 0.318 & 0.292 & 0.000 & 0.028 \\
msmarco\_v2.1\_doc\_47\_1430382251 & 0.286 & 0.273 & 0.000 & 0.114 \\
msmarco\_v2.1\_doc\_48\_1289995263 & 0.107 & 0.103 & 0.000 & 0.000 \\
msmarco\_v2.1\_doc\_48\_273997000 & 0.000 & 0.000 & 0.000 & 0.057 \\
msmarco\_v2.1\_doc\_48\_515083157 & 0.087 & 0.083 & 0.083 & 0.219 \\
msmarco\_v2.1\_doc\_48\_515287844 & 0.077 & 0.000 & 0.000 & 0.190 \\
msmarco\_v2.1\_doc\_48\_730773621 & 0.471 & 0.444 & 0.000 & 0.111 \\
msmarco\_v2.1\_doc\_52\_1666095905 & 0.088 & 0.028 & 0.102 & 0.296 \\
msmarco\_v2.1\_doc\_54\_1547893878 & 0.000 & 0.000 & 0.000 & 0.000 \\
msmarco\_v2.1\_doc\_54\_304836636 & 0.114 & 0.109 & 0.000 & 0.009 \\
msmarco\_v2.1\_doc\_55\_248024230 & 0.038 & 0.037 & 0.000 & 0.000 \\
msmarco\_v2.1\_doc\_57\_933363597 & 0.308 & 0.286 & 0.000 & 0.214 \\
msmarco\_v2.1\_doc\_58\_748655897 & 0.000 & 0.000 & 0.000 & 0.029 \\
\bottomrule
\end{tabular}

\caption{Results of our submitted run for question generation (\textit{qgen\_score original} represents the scores of the original evaluation while \textit{qgen\_score updated} are updated evaluation results the DRAGUN organizers sent after slightly changing the evaluation setup for task 1) as well as contradictory score (the lower, the better) and supportive score (the higher, the better) for generated report.}
\end{table*}

\begin{figure}[htbp]
    \centering
    \includegraphics[width=\linewidth]{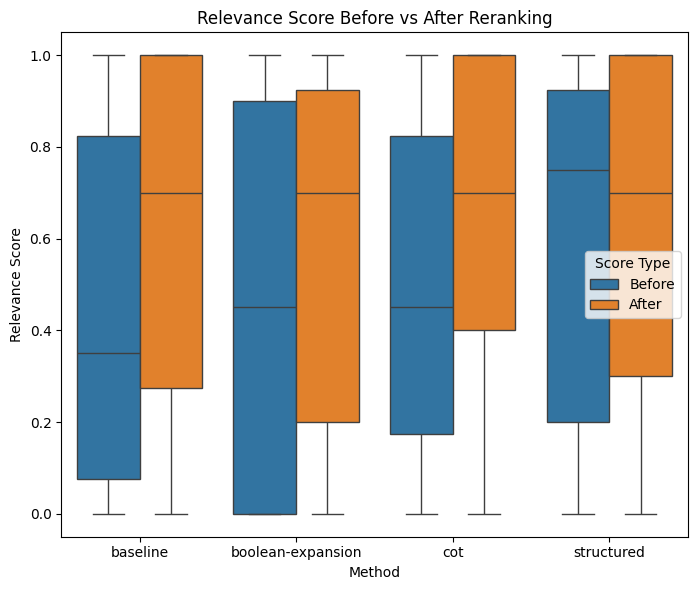}
    \caption{Relevance scores for the first 20 example TREC questions before and after re-ranking across all four query expansion methods.}
    \label{fig:baseline_relevance}
\end{figure}

\begin{figure}[htbp]
    \centering
    \includegraphics[width=\linewidth]{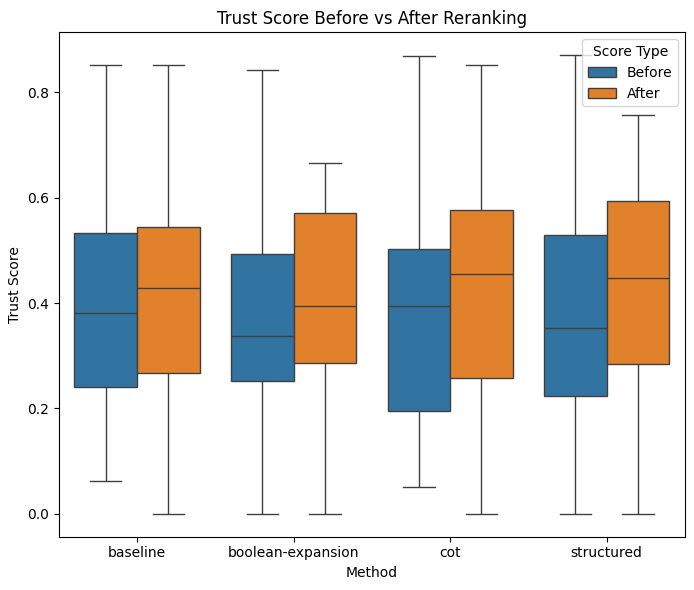}
    \caption{Domain trustworthiness scores for the first 20 example TREC questions before and after re-ranking across all four query expansion methods.}
    \label{fig:baseline_trust}
\end{figure}

\begin{figure}[htbp]
    \centering
    \includegraphics[width=\linewidth]{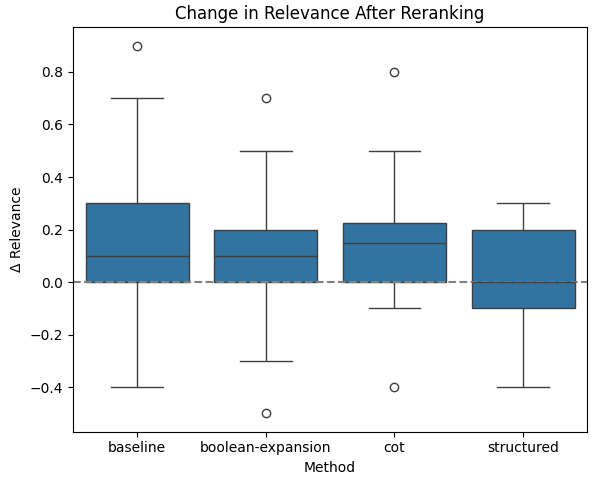}
    \caption{Change in relevance scores after re-ranking for all four query expansion methods.}
    \label{fig:delta_relevance}
\end{figure}

\begin{figure}[htbp]
    \centering
    \includegraphics[width=\linewidth]{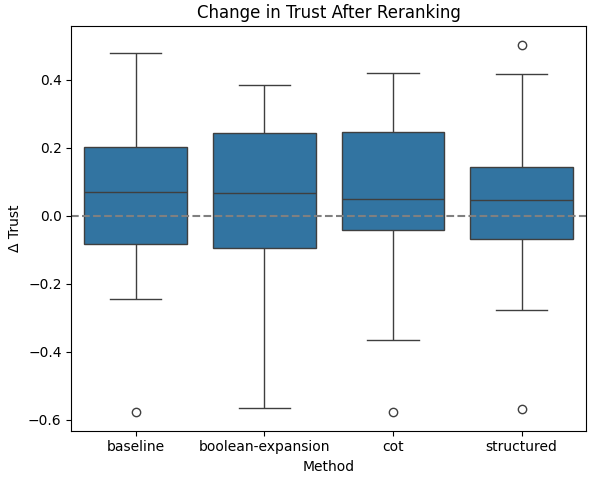}
    \caption{Change in domain trustworthiness scores after re-ranking for all four query expansion methods.}
    \label{fig:delta_trust}
\end{figure}

\section{Discussion}



Our results show that Chain-of-Thought based query expansion combined with monoT5 re-ranking yielded the best trade-off between relevance and domain trust scores for Task 2. This supports the use of reasoning-oriented expansion as a way to expose missing facets in news-related information needs. In contrast, question generation and report writing lagged behind, with low supportive scores, high topic-level variance, and inconsistencies in LLM-based scoring. These issues were amplified by our reliance on LLM components for internal evaluation and relevance judgments, which introduces calibration and alignment uncertainty. 

A key limitation of our study is that we submitted only a single run and were unable to integrate the planned iterative correction loop or more advanced question-quality filtering into the final pipeline. As a result, parts of the system remained prototype-level and not tuned against the official evaluation setup. Taken together, the findings suggest that future work should focus on stabilizing question generation, improving LLM-based filtering and scoring, and strengthening the coupling between retrieval, trust signals, and downstream report structure.

\section{Conclusion}



In this paper, we presented UR\_Trecking, a retrieval-augmented LLM-based pipeline for critical question generation and trustworthiness reporting in the DRAGUN Track. Our main contribution is an empirical analysis of Chain-of-Thought query expansion plus neural re-ranking augmented with domain-level trust signals, together with an end-to-end implementation for news trustworthiness support on MS~MARCO~V2.1 segmented. The evaluation indicates gains in retrieval relevance and domain trustworthiness, but only moderate alignment of generated questions and reports with assessor rubrics. Next steps are to complete and stabilize the pipeline by integrating iterative correction, making LLM-based scoring more robust, and explicitly optimizing for rubric coverage. In the longer term, user studies will be needed to test whether these technical improvements translate into better trust judgments in real-world news consumption.


\bibliographystyle{ACM-Reference-Format}
\bibliography{sample-base}



\end{document}